\documentclass[12pt]{article}
\usepackage{amssymb, amsfonts, amsmath, latexsym}
\def\ra{\rangle}
\def\la{\langle}

\begin{document}

\title{The tripartite separability of density matrices of graphs}
\author{Zhen Wang and Zhixi Wang\\
{\small Department of Mathematics}\\
{\small Capital  Normal University, Beijing 100037, China}\\
{\small wangzhen061213@sina.com,\  wangzhx@mail.cnu.edu.cn}}
\date{}
\maketitle

\begin{abstract} The density matrix of a graph is the combinatorial laplacian
matrix of a graph normalized to have unit trace. In this paper we
generalize the entanglement properties of mixed density matrices
from combinatorial laplacian matrices of graphs discussed in
Braunstein {\it et al.} Annals of Combinatorics, {\bf 10}(2006)291
to tripartite states. Then we proved that the degree condition
defined in Braunstein {\it et al.} Phys. Rev. A {\bf 73},
(2006)012320 is sufficient and necessary for the tripartite
separability of the density matrix of a nearest point graph.
\end{abstract}

\section{Introduction}
\quad Quantum entanglement is one of the most striking features of
the quantum formalism$^{\tiny\cite{peres1}}$. Moreover, quantum
entangled states may be used as basic resources in quantum
information processing and communication, such as quantum
cryptography$^{\tiny\cite{ekert}}$, quantum
parallelism$^{\tiny\cite{deutsch}}$, quantum dense
coding$^{\tiny\cite{bennett1,mattle}}$ and quantum
teleportation$^{\tiny\cite{bennett2,bouwmeester}}$. So testing
whether a given state of a composite quantum system is separable or
entangled is in general very important.

Recently, normalized laplacian matrices of graphs considered as
density matrices have been studied in quantum mechanics. One can
recall the definition of density matrices of graphs from
\cite{sam1}. Ali Saif M. Hassan and Pramod Joag$^{\tiny\cite{Ali}}$
studied the related issues like classification of pure and mixed
states, von Neumann entropy, separability of multipartite quantum
states and quantum operations in terms of the graphs associated with
quantum states. Chai Wah Wu$^{\tiny\cite{chai}}$ showed that the
Peres-Horodecki positive partial transpose condition is necessary
and sufficient for separability in $C^2\otimes C^q$. Braunstein {\it
et al.}$^{\tiny\cite{sam2}}$ proved that the degree condition is
necessary for separability of density matrices of any graph and is
sufficient for separability of density matrices of nearest point
graphs and perfect matching graphs. Ali Saif M. Hassan and Pramod
Joag shows that the degree condition is also necessary and
sufficient condition for the separability of $m$-partite pure
quantum states living in a real or complex Hilbert space in
$\cite{Ali2}$. Hildebrand {\it et al.}$^{\tiny\cite{roland}}$
testified that the degree condition is equivalent to the
PPT-criterion. They also considered the concurrence of density
matrices of graphs and pointed out that there are examples on four
vertices whose concurrence is a rational number.

The paper is divided into three sections. In section 2, we recall
the definition of the density matrices of a graph and define the
tensor product of three graphs, reconsider the tripartite
entanglement properties of the density matrices of graphs introduced
in \cite{sam1}. In section 3, we define partially transposed graph
at first and then shows that the degree condition introduced in
\cite{sam2} is also sufficient and necessary condition for the
tripartite mixed state of the density matrices of nearest point
graphs.

\section{The tripartite entanglement properties of the
density matrices of graphs}

\quad Recall that from \cite{sam1} a {\it graph} $G=(V(G),\ E(G))$
is defined as: $V(G)=\{v_1,\ v_2,\ \cdots,\ v_n\}$ is a non-empty
and finite set called {\it vertices}; $E(G)=\{\{v_i,\ v_j\}:\ v_i,\
v_j\in V\}$ is a non-empty set of unordered pairs of vertices called
{\it edges}. An edge of the form $\{v_i,\ v_i\}$ is called as a {\it
loop}. We assume that $E(G)$ does not contain any loops. A graph $G$
is said to be on $n$ vertices if $|V(G)|=n$. The {\it adjacency
matrix} of a graph $G$ on $n$ vertices is an $n\times n$ matrix,
denoted by $M(G)$, with lines labeled by the vertices of $G$ and
$ij$-th entry defined as:
$$[M(G)]_{i,j}=\left\{
\begin{array}{ll}
1, & \hbox{if $(v_{i},\ v_{j})\in E(G)$;}\\
0, & \hbox{if $(v_{i},\ v_{j})\notin E(G)$.}
\end{array}
\right.$$

If $\{v_i,\ v_j\}\in E(G)$ two distinct vertices $v_i$ and $v_j$ are
said to be {\it adjacent}. The {\it degree} of a vertex $v_i\in
V(G)$ is the number of edges adjacent to $v_i$, we denote it as
$d_G(v_i)$. $d_G=\displaystyle\sum_{i=1}^nd_G(v_i)$ is called as the
{\it degree sum}. Notice that $d_G=2|E(G)|.$ The {\it degree matrix}
of $G$ is an $n\times n$ matrix, denoted as $\Delta(G)$, with
$ij$-th entry defined as:
$$[\Delta(G)]_{i,\ j}=\left\{
\begin{array}{ll}
d_{G}(v_{i}), & \hbox{if $i=j$;\ }\\
0, & \hbox{if $i\neq j$.\ }
\end{array}
\right.$$

The {\it combinatorial laplacian matrix} of a graph $G$ is the
symmetric positive semidefinite matrix
$$L(G)=\Delta(G)-M(G).$$

The density matrix of $G$ of a graph $G$ is the matrix
$$\rho(G)=\frac{1}{d_{G}}L(G).$$

Recall that a graph is called {\it complete}$^{\tiny\cite{gtm207}}$
if every pair of vertices are adjacent, and the {\it complete graph}
on $n$ vertices is denoted by $K_n$. Obviously,
$\rho(K_n)=\frac{1}{n(n-1)}(nI_n-J_n),$ where $I_n$ and $J_n$ is the
$n\times n$ identity matrix and the $n\times n$ all-ones matrix,
respectively. A {\it star graph} on $n$ vertices $\alpha_1,\
\alpha_2,\ \cdots,\ \alpha_n$, denoted by $K_{1,n-1}$, is the graph
whose set of edges is $\{\{\alpha_1,\ \alpha_i\}:\ i=2,\ 3,\
\cdots,\ n\}$, we have
$$\rho(K_{1,n-1})
=\frac{1}{2(n-1)} \left(
\begin{array}{ccccc}
n-1&-1&-1&\cdots&-1\\[3mm]
-1&1&&&\\[3mm]
-1&&1&&\\[3mm]
\vdots&&&\ddots&\\[3mm]
-1&&&&1
\end{array}
\right).$$

Let $G$ be a graph which has only a edge. Then the density matrix of
$G$ is pure. The density matrix of a graph is a uniform mixture of
pure density matrices, that is, for a graph $G$ on $n$ vertices
$v_1,\ v_2,\ \cdots,\ v_n,$ having $s$ edges $\{v_{i_1},\
v_{j_1}\},\ \{v_{i_2},\ v_{j_2}\},\ \cdots,\ \{v_{i_s},\ v_{j_s}\},$
where $1\leq i_1,\ j_1,\ i_2,\ j_2,\ \cdots,\ i_k,\ j_k\leq n,$
$$\rho(G)=\displaystyle\frac{1}{s}\sum_{k=1}^{s}\rho(H_{i_kj_k}),$$
here $H_{i_kj_k}$ is the factor of $G$ such that
$$[M(H_{i_kj_k})]_{u,\ w}=\left\{
\begin{array}{ll}
1, & \hbox{if}\ u=i_k\ \hbox{and}\ w=j_k\ \hbox{or}\
w=i_k\ \hbox{and}\ u=j_k;\\
0, & \hbox{otherwise.}
\end{array}
\right.$$ It is obvious that $\rho(H_{i_kj_k})$ is pure.

Before we discuss the tripartite entanglement properties of the
density matrices of graphs we will at first recall briefly the
definition of the tripartite separability:

{\bf Definition 1}\quad The state $\rho$ acting on ${\cal H}={\cal
H_A}\otimes{\cal H_B}\otimes{\cal H_C}$ is called {\it tripartite
separability} if it can be written in the form
$$\rho=\displaystyle\sum_i p_i\rho_A^i\otimes\rho_B^i\otimes\rho_C^i,$$
where $\rho_A^i=|\alpha_A^i\ra\la\alpha_A^i|,\
\rho_B^i=|\beta_B^i\ra\la\beta_B^i|,\
\rho_C^i=|\gamma_C^i\ra\la\gamma_C^i|,\ \displaystyle\sum_i p_i=1,\
p_i\geq0$ and $|\alpha_A^i\ra$,\ $|\beta_B^i\ra$,\ $|\gamma_C^i\ra$ are
normalized pure states of subsystems $A,\ B$ and $C$, respectively.
Otherwise, the state is called {\it entangled}.

Now we define the tensor product of three graphs. The {\it tensor
product of graphs} $G_A,\ G_B,\ G_C$, denoted by $G_A\otimes
G_B\otimes G_C$, is the graph whose adjacency matrix is
$M(G_A\otimes G_B\otimes G_C)=M(G_A)\otimes M(G_B)\otimes M(G_C).$
Whenever we consider a graph $G_A\otimes G_B\otimes G_C$, where
$G_A$ is on $m$ vertices, $G_B$ is on $p$ vertices and $G_C$ is on
$q$ vertices, the tripartite separability of $\rho(G_A\otimes
G_B\otimes G_C)$ is described with respect to the Hilbert space
${\cal H}_A\otimes {\cal H}_B\otimes {\cal H}_C$, where ${\cal H}_A$
is the space spanned by the orthonormal basis $\{|u_1\ra,\
|u_{2}\ra,\ \cdots,\ |u_m\ra\}$ associated to $V(G_A)$, ${\cal H}_B$
is the space spanned by the orthonormal basis $\{|v_1\ra,\
|v_{2}\ra,\ \cdots,\ |v_p\ra\}$ associated to $V(G_B)$ and ${\cal
H}_C$ is the space spanned by the orthonormal basis $\{|w_1\ra,\
|w_2\ra,\ \cdots,\ |w_q\ra\}$ associated to $V(G_C)$. The vertices
of $G_A\otimes G_B\otimes G_C$ are taken as $\{u_iv_jw_k,\ 1\leq
i\leq m,\ 1\leq j\leq p,\ 1\leq k\leq q\}.$ We associate
$|u_i\ra|v_j\ra|w_k\ra$ to $u_iv_jw_k,$ where $1\leq i\leq m,\ 1\leq
j\leq p,\ 1\leq k\leq q.$ In conjunction with this, whenever we talk
about tripartite separability of any graph $G$ on $n$ vertices,
$|\alpha_1\ra,\ |\alpha_2\ra,\ \cdots,\ |\alpha_n\ra,$ we consider
it in the space $C^m\otimes C^p\otimes C^q$, where $n=mpq$. The
vectors $|\alpha_1\ra,\ |\alpha_2\ra,\ \cdots,\ |\alpha_n\ra$ are
taken as follows: $|\alpha_1\ra=|u_1\ra|v_1\ra|w_1\ra,\
|\alpha_2\ra=|u_1\ra|v_1\ra|w_2\ra,\ \cdots,\
|\alpha_n\ra=|u_m\ra|v_p\ra|w_q\ra.$

To investigate the tripartite entanglement properties of the density
matrices of graphs it is necessary to recall the well known positive
partial transposition criterion (i.e. Peres criterion). It makes use
of the notion of {\it partial transpose} of a density matrix. Here
we will only recall the Peres criterion for the tripartite states.
Consider a $n\times n$ matrix $\rho_{ABC}$ acting on $C_A^m\otimes
C_B^p\otimes C_C^q$, where $n=mpq$. The partial transpose of
$\rho_{ABC}$ with respect to the systems $A,\ B,\ C$ are the
matrices $\rho^{T_{A}}_{ABC},\ \rho^{T_{B}}_{ABC},\
\rho^{T_{C}}_{ABC}$, respectively, and with $(i,\ j,\ k;\ i^\prime,\
j^\prime,\ k^\prime)$-th entry defined as follows:
$$\begin{array}{rl}
&[\rho^{T_{A}}_{ABC}]_{i,\ j,\ k;\ i^{\prime},\ j^{\prime},\
k^{\prime}}=
\la u_{i^{\prime}}v_{j}w_{k}|\rho_{ABC}|u_{i}v_{j^{\prime}}w_{k^{\prime}}\ra,\\[3mm]
&[\rho^{T_{B}}_{ABC}]_{i,\ j,\ k;\ i^{\prime},\ j^{\prime},\
k^{\prime}}=
\la u_{i}v_{j^{\prime}}w_{k}|\rho_{ABC}|u_{i^{\prime}}v_{j}w_{k^{\prime}}\ra,\\[3mm]
&[\rho^{T_{C}}_{ABC}]_{i,\ j,\ k;\ i^{\prime},\ j^{\prime},\
k^{\prime}}= \la
u_{i}v_{j}w_{k^{\prime}}|\rho_{ABC}|u_{i^{\prime}}v_{j^{\prime}}w_{k}\ra,
\end{array}$$
where $1 \leq i,\ i^{\prime} \leq m;\ 1\leq j,\ j^{\prime} \leq p\
\hbox{and}\ 1 \leq k,\ k^{\prime} \leq q.$

For separability of $\rho_{ABC}$ we have the following criterion:

{\bf Peres criterion}$^{\tiny\cite{peres2}}$ \ If $\rho$ is a
separable density matrix acting on $C^m\otimes C^p\otimes C^q$, then
$\rho^{T_A},\ \rho^{T_B},\ \rho^{T_C}$ are positive semidefinite.

{\bf Lemma 1}\quad The density matrix of the tensor product of three
graphs is tripartite separable.

{\bf Proof.}\quad Let $G_1$ be a graph on $n$ vertices, $u_1,\ u_2,\
\cdots,\ u_n$, and $m$ edges, $\{u_{c_1}$,\ $u_{d_1}\}$,\ $\cdots,
\{u_{c_m}, u_{d_m}\},$ $1\leq c_1,\ d_1,\ \cdots,\ c_m,\ d_m\leq n.$
Let $G_2$ be a graph on $k$ vertices, $v_1,\ v_2,\ \cdots,\ v_k$,
and $e$ edges,
$\{v_{i_1},\ v_{j_1}\},\ \cdots,\ \{v_{i_e},\
v_{j_e}\},\ 1\leq i_1,\ j_1,\ \cdots,\ i_e,\ j_e\leq k.$
Let $G_3$
be a graph on $l$ vertices, $w_1,\ w_2,\ \cdots,\ w_l$, and $f$
edges,
$\{w_{r_1},\ w_{s_1}\},\ \cdots,\ \{w_{r_f},\ w_{s_f}\},\
1\leq r_1,\ s_1,\ \cdots,\ r_f,\ s_f\leq l.$
Then
$$\rho(G_1)=\frac{1}{m}\sum_{p=1}^m\rho(H_{c_pd_p}),\
\rho(G_2)=\frac{1}{e}\sum_{q=1}^e\rho(L_{i_qj_q}),\
\rho(G_3)=\frac{1}{f}\sum_{t=1}^f\rho(Q_{r_ts_t}).$$
Therefore
$$\begin{array}{rl}
&\hskip-1cm \rho(G_1\otimes G_2\otimes G_3)\\[3mm]
&=\displaystyle\frac{1}{d_{G_1\otimes G_2\otimes
G_3}}[\Delta(G_1\otimes G_2\otimes G_3)-M(G_1\otimes
G_2\otimes G_3)]
\end{array}$$

$$\begin{array}{rl}
&=\displaystyle\frac{1}{d_{G_1\otimes G_2\otimes
G_3}}\sum_{p=1}^m\sum_{q=1}^e\sum_{t=1}^f[\Delta(H_{c_pd_p}\otimes
L_{i_qj_q}\otimes Q_{r_ts_t})-M(H_{c_pd_p}\otimes
L_{i_qj_q}\otimes Q_{r_ts_t} )]\\
&=\displaystyle\frac{1}{d_{G_1\otimes G_2\otimes
G_3}}\sum_{p=1}^m\sum_{q=1}^e\sum_{t=1}^f8 \rho(H_{c_pd_p}\otimes
L_{i_qj_q}\otimes Q_{r_ts_t})\\
&=\displaystyle\frac{1}{mef}\sum_{p=1}^m\sum_{q=1}^e\sum_{t=1}^f
\rho(H_{c_pd_p}\otimes L_{i_qj_q}\otimes
Q_{r_ts_t})\\
&=\displaystyle\frac{1}{mef}\sum_{p=1}^m\sum_{q=1}^e\sum_{t=1}^f
\frac{1}{8}[\Delta(H_{c_pd_p})\otimes\Delta(L_{i_qj_q})\otimes
\Delta(Q_{r_ts_t})-M(H_{c_pd_p})\otimes M(L_{i_qj_q})\otimes
M(Q_{r_ts_t})]\\
&=\displaystyle\frac{1}{mef}\sum_{p=1}^m\sum_{q=1}^e\sum_{t=1}^f
\frac{1}{4}[\rho(H_{c_pd_p})\otimes\rho(L_{i_qj_q})\otimes
\rho(Q_{r_ts_t})\\
&+\rho_+(H_{c_pd_p})\otimes\rho(L_{i_qj_q})\otimes
\rho_+(Q_{r_ts_t}) +\rho(H_{c_pd_p})\otimes\rho_+(L_{i_qj_q})\otimes
\rho_+(Q_{r_ts_t})\\[3mm]
&+\rho_+(H_{c_pd_p})\otimes\rho_+(L_{i_qj_q})\otimes
\rho(Q_{r_ts_t})],
\end{array}$$
where
$$\rho_+(H_{c_pd_p})\stackrel{\rm
def}{=}\Delta(H_{c_pd_p})-\rho(H_{c_pd_p})
=\frac{1}{2}\big(\Delta(H_{c_pd_p})+M(H_{c_pd_p})\big),$$
$$\rho_+(L_{i_qj_q})\stackrel{\rm
def}{=}\Delta(L_{i_qj_q})-\rho(L_{i_qj_q})
=\frac{1}{2}\big(\Delta(L_{i_qj_q})+M(L_{i_qj_q})\big),$$
$$\rho_+(Q_{r_ts_t})\stackrel{\rm
def}{=}\Delta(Q_{r_ts_t})-\rho(Q_{r_ts_t})
=\frac{1}{2}\big(\Delta(Q_{r_ts_t})+M(Q_{r_ts_t})\big),$$ the fourth
equality follows from $d_{G_1\otimes G_2\otimes G_3}=8mef$ and the
fifth equality follows from the definition of tensor products of
graphs.

Notice that
$\rho_+(H_{c_pd_p}),\ \rho_+(L_{i_qj_q}),\
\rho_+(Q_{r_ts_t})$ are all density matrices.
Let
$$\rho_+(G_1)=\frac{1}{m}\sum_{p=1}^m
\rho_+(H_{c_pd_p}),\quad \rho_+(G_2)=\frac{1}{e}\sum_{q=1}^e
\rho_+(L_{i_qj_q}),\quad \rho_+(G_3)=\frac{1}{f}\sum_{t=1}^f
\rho_+(Q_{r_ts_t}).$$
Then
$$\begin{array}{rl}
\rho(G_1\otimes G_2\otimes G_3)
=&\frac{1}{4}[\rho(G_1)\otimes\rho(G_2)
\otimes\rho(G_3)+\rho_+(G_1)\otimes\rho(G_2)
\otimes\rho_+(G_3)\\[3mm]
&+\rho(G_1)\otimes\rho_+(G_2)\otimes\rho_+(G_3)
+\rho_+(G_1)\otimes\rho_+(G_2)\otimes\rho(G_3)].
\end{array}$$
So we have that $\rho(G)$ is tripartite separable. $\Box$

{\bf Remark}\quad We associate to the vertices $\alpha_1,\
\alpha_2,\ \cdots, \alpha_n$ of a graph $G$ an orthonormal basis
$\{|\alpha_1\ra,\ |\alpha_2\ra,$ $\cdots,\ |\alpha_n\ra\}.$ In terms
of this basis, the $uw$-th elements of the matrices
$\rho(H_{c_pd_p})$ and $\rho_+(H_{c_pd_p})$ are given by
$\la\alpha_u|\rho(H_{c_pd_p})|\alpha_w\ra$ and
$\la\alpha_u|\rho_+(H_{c_pd_p})|\alpha_w\ra$, respectively. In this
basis we have
$$\rho(H_{c_pd_p})=
P[\frac{1}{\sqrt{2}}(|\alpha_{c_p}\ra-|\alpha_{d_p}\ra)],\
\rho_+(H_{c_pd_p})=
P[\frac{1}{\sqrt{2}}(|\alpha_{c_p}\ra+|\alpha_{d_p}\ra)].$$

{\bf Lemma 2}\quad The matrix $\sigma=\frac{1}{4}P[\frac{1}
{\sqrt{2}}(|ijk\ra-|rst\ra )]
+\frac{1}{4}P[\frac{1}{\sqrt{2}}(|ijt\ra -|rsk\ra )]+\frac{1}{4}
P[\frac{1}{\sqrt{2}}(|isk\ra -|rjt\ra )]
+\frac{1}{4}P[\frac{1}{\sqrt{2}}(|rjk\ra -|ist\ra )]$ is a density
matrix and tripartite separable.

{\bf Proof. }\quad Since the project operator is semipositive,
$\sigma$ is semipositive. By computing one can get $tr(\sigma)=1$,
so $\sigma$ is a density matrix. Let
$$|u^{\pm}\ra=\frac{1}{\sqrt{2}}(|i\ra\pm|r\ra),\ \
|v^{\pm}\ra=\frac{1}{\sqrt{2}}(|j\ra\pm|s\ra),\ \
|w^{\pm}\ra=\frac{1}{\sqrt{2}}(|k\ra\pm|t\ra).$$ We obtain
$$\sigma=\frac{1}{4}
P[|u^{+}\ra|v^{-}\ra|w^{+}\ra]
+\frac{1}{4}P[|u^{+}\ra|v^{+}\ra|w^{-}\ra]
+\frac{1}{4}P[|u^{-}\ra|v^{-}\ra|w^{-}\ra]
+\frac{1}{4}P[|u^{-}\ra|v^{+}\ra|w^{+}\ra],$$ thus $\sigma$ is
tripartite separable. $\Box$

{\bf Lemma 3}\quad For any $n=mpq$, the density matrix $\rho(K_n)$
is tripartite separable in $C^m\otimes C^p\otimes C^q.$

{\bf Proof.}\quad Since $M(K_n)=J_n-I_n$, where $J_n$ is the
$n\times n$ all-ones matrix and $I_n$ is the $n\times n$ identity
matrix, whenever there is an edge $\{u_{i}v_{j}w_{k},\
u_{r}v_{s}w_{t}\}$, there must be entangled edges
$\{u_{r}v_{j}w_{k},\ u_{i}v_{s}w_{t}\},$ $\{u_{i}v_{s}w_{k},\
u_{r}v_{j}w_{t}\}$ and $\{u_{i}v_{j}w_{t},\ u_{r}v_{s}w_{k}\}.$ The
result follows from Lemma 2. $\Box$

{\bf Lemma 4}\quad The complete graph on $n>1$ vertices is not a
tensor product of three graphs.

{\bf Proof.}\quad It is obvious that $K_n$ is not a tensor product
of three graphs if $n$ is a prime or a product of two primes. Thus
we can assume that $n$ is a product of three or more primes. Let
$n=mpq,\ m,\ p,\ q>1.$ Suppose that there exist three graphs $G_1,\
G_2$ and $G_3$ on $m,\ p$ and $q$ vertices, respectively, such that
$K_{mpq}=G_1\otimes G_2\otimes G_3.$ Let $|E(G_1)|=r,\ |E(G_2)|=s,\
|E(G_3)|=t.$ Then, by the degree sum formula, $2r\leq m(m-1),\
2s\leq p(p-1),\ 2t\leq q(q-1).$ Hence
$$2r\cdot 2s\cdot 2t\leq
mpq(m-1)(p-1)(q-1)=mpq(mpq-mp-mq-pq+m+p+q-1).$$
Now, observe that
$$|V(G_1\otimes G_2\otimes G_3)|=mpq,\quad |E(G_1\otimes G_2\otimes
G_3)|=4rst.$$
Therefore,
$$G_1\otimes G_2\otimes G_3=K_{mpq} \iff  mpq(mpq-1)=2\cdot 4rst,$$
so
$$ mpq(mpq-1)=8rst\leq mpq(mpq-mp-mq-pq+m+p+q-1).$$
It follows that $mp+mq+pq-m-p-q\leq 0$,
that is
$m(p-1)+q(m-1)+p(q-1)\leq 0.$
As $m,p,q\geq 1$ we get
$m(p-1)+q(m-1)+p(q-1)=0$.
It yields that $m=p=q=1.$ $\Box$

{\bf Theorem 1}\quad Given a graph $G_1\otimes G_2\otimes G_3$, the
density matrix $\rho(G_1\otimes G_2\otimes G_3)$ is tripartite
separable. However if a density matrix $\rho(L)$ is tripartite
separable it does not necessarily mean that $L=L_1\otimes L_2\otimes
L_3,$ for some graphs $L_1,\ L_2$ and $L_3.$

{\bf Proof.}\quad The result follows from Lemmas 1, 3 and 4. $\Box$

{\bf Theorem 2}\quad The density matrix $\rho(K_{1,\ n-1})$ is
tripartite entangled for $n=mpq\geq 8.$

{\bf Proof.}\quad Consider a graph $G=K_{1,\ n-1}$ on $n=mpq$
vertices,
$|\alpha_1\ra,\ |\alpha_2\ra,\ \cdots,\ |\alpha_n\ra.$
Then
$$\rho(G)=\frac{1}{n-1}\sum_{k=2}^n\rho(H_{1k})=
\frac{1}{n-1}\sum_{k=2}^nP[\frac{1}{\sqrt{2}}(|\alpha_1\ra-|\alpha_n\ra)].$$
We are going to examine tripartite separability of $\rho(G)$ in
$C^m_A\otimes C^p_B\otimes C^q_C,$
where $C^m_A,\ C^p_B$ and $C^q_C$
are associated to three quantum systems ${\cal H}_A,\ {\cal H}_B$
and ${\cal H}_C,$ respectively.
Let
$\{|u_1\ra,\ |u_2\ra,\ \cdots,\
|u_m\ra\}$, $\{|v_1\ra,\ |v_2\ra,\ \cdots,$ $|v_p\ra\}$ and
$\{|w_1\ra,\ |w_2\ra,\ \cdots,\ |w_q\ra\}$ be orthonormal basis of
$C^m_A,\ C^p_B$ and $C^q_C,$ respectively.
So,
$$\rho(G)=\frac{1}{n-1}\sum_{k=2}^{n}P[\frac{1}
{\sqrt{2}}(|u_1v_1w_1\ra-|u_{r_k}v_{s_k}w_{t_k}\ra)],$$ where
$k=(r_k-1)pq+(s_k-1)+t_k,\ 1\leq r_k\leq m,\ 1\leq s_k\leq p,\ 1\leq
t_k\leq q.$
Hence
$$\begin{array}{rl}
\rho(G)=&\displaystyle\frac{1}{n-1}\Big\{\sum_{i=2}^{m}P[\frac{1}
{\sqrt{2}}(|u_1\ra-|u_i\ra)|v_1\ra|w_1\ra]+\sum_{j=2}^{p}P[
|u_1\ra\frac{1}{\sqrt{2}}(|v_1\ra-|v_j\ra)|w_1\ra]\\[3mm]
&+\displaystyle\sum_{k=2}^{q}P[|u_1\ra|v_1\ra\frac{1}{\sqrt{2}}
(|w_1\ra-|w_k\ra)] +\sum_{i=2}^{m}\sum_{j=2}^{p}P[\frac{1}{\sqrt{2}}
(|u_1v_1w_1\ra-|u_iv_jw_1\ra)]\\[3mm]
&+\displaystyle\sum_{j=2}^{p}\sum_{k=2}^{q}P[\frac{1}{\sqrt{2}}
(|u_1v_1w_1\ra-|u_1v_jw_k\ra)]
+\sum_{i=2}^{m}\sum_{k=2}^{q}P[\frac{1}{\sqrt{2}}
(|u_1v_1w_1\ra-|u_iv_1w_k\ra)]\\[3mm]
&+\displaystyle\sum_{i=2}^{m}\sum_{j=2}^{p}\sum_{k=2}^{q}
P[\frac{1}{\sqrt{2}} (|u_1v_1w_1\ra-|u_iv_jw_k\ra)]\Big\}.
\end{array}$$
Consider now the following projectors:
$$P=|u_1\ra\la u_1|+|u_2\ra\la
u_2|,\quad Q=|v_1\ra\la v_1|+|v_2\ra\la v_2|\  {\rm \ and\ }\ R=|w_1\ra\la
w_1|+|w_2\ra\la w_2|.$$
Then
$$\begin{array}{rl}
&\hskip-2.6cm(P\otimes Q\otimes R)\rho(G)(P\otimes Q\otimes R)\\
=&\frac{1}{n-1}\Big\{\frac{n-8}{2}P[|u_1v_1w_1\ra]+
P[\frac{1}{\sqrt{2}}(|u_1v_1w_1\ra-|u_1v_1w_2\ra)]\\[3mm]
&+P[\frac{1}{\sqrt{2}}(|u_1v_1w_1\ra-|u_1v_2w_1\ra)]
+P[\frac{1}{\sqrt{2}}(|u_1v_1w_1\ra-|u_2v_1w_1\ra)]\\[3mm]
&+P[\frac{1}{\sqrt{2}}(|u_1v_1w_1\ra-|u_1v_2w_2\ra)]
+P[\frac{1}{\sqrt{2}}(|u_1v_1w_1\ra-|u_2v_1w_2\ra)]\\[3mm]
&+P[\frac{1}{\sqrt{2}}(|u_1v_1w_1\ra-|u_2v_2w_1\ra)]
+P[\frac{1}{\sqrt{2}}(|u_1v_1w_1\ra-|u_2v_2w_2\ra)]\Big\}.
\end{array}$$
In the basis
$$\{|u_1v_1w_1\ra, |u_1v_1w_2\ra, |u_1v_2w_1\ra,
|u_1v_2w_2\ra, |u_2v_1w_1\ra,\ |u_2v_1w_2\ra,\ |u_2v_2w_1\ra,
|u_2v_2w_2\ra\},$$
we have
$$[(P\otimes Q\otimes R)\rho(G)(P\otimes Q\otimes R)]^{T_{A}}
=\frac{1}{n-1} \left(
\begin{array}{cccccccccccc}
\frac{n-1}{2}&-\frac{1}{2}&-\frac{1}{2}&-\frac{1}{2}&-\frac{1}{2}&0&0&0\\[3mm]
-\frac{1}{2}&\frac{1}{2}&0&0&-\frac{1}{2}&0&0&0\\[3mm]
-\frac{1}{2}&0&\frac{1}{2}&0&-\frac{1}{2}&0&0&0\\[3mm]
-\frac{1}{2}&0&0&\frac{1}{2}&-\frac{1}{2}&0&0&0\\[3mm]
-\frac{1}{2}&-\frac{1}{2}&-\frac{1}{2}&-\frac{1}{2}&\frac{1}{2}&0&0&0\\[3mm]
0&0&0&0&0&\frac{1}{2}&0&0\\[3mm]
0&0&0&0&0&0&\frac{1}{2}&0\\[3mm]
0&0&0&0&0&0&0&\frac{1}{2}
\end{array}
\right).$$

The eigenpolynomial of the above matrix is
$$\Big(\lambda-\frac{1}{2(n-1)}\Big)^{5}\Big(\lambda^{3}
-\frac{n+1}{2(n-1)}\lambda^{2}
+\frac{n-4}{2(n-1)^{2}}\lambda+\frac{n+4}{4(n-1)^{3}}\Big),$$
so the
eigenvalues of the matrix are $\frac{1}{2(n-1)}$ (with multiplicity
5) and the roots of the polynomial
$\lambda^{3}-\frac{n+1}{2(n-1)}\lambda^{2}
+\frac{n-4}{2(n-1)^{2}}\lambda+\frac{n+4}{4(n-1)^{3}}.$ Let the
roots of this polynomial of degree three be $\lambda_1,\ \lambda_2$
and $\lambda_3$. Then
$\lambda_{1}\lambda_{2}\lambda_{3}=-\frac{n+4}{4(n-1)^{3}}<0,$ so
one of the three roots must be negative, i.e., there must be a
negative eigenvalue of the above matrix. Hence, by Peres criterion,
the matrix $(P\otimes Q\otimes R)\rho(G)(P\otimes Q\otimes R)$ is
tripartite entangled and then $\rho(G)$ is tripartite entangled.
$\Box$

\section{A sufficient and necessary condition of tripartite separability}

\indent {\bf Definition 2}\quad {\it Partially transposed graph}
$G^{\Gamma_A}=(V,\ E^\prime)$, (i.e. the partial transpose of a
graph $G=(V,E)$ with respect to ${\cal H}_A$) is the graph such that
$$\{u_iv_jw_k,\ u_rv_sw_t\}\in E^\prime \ \hbox{if and only if}\ \{u_rv_jw_k,\
u_iv_sw_t\}\in E.$$ Partially transposed graphs $G^{\Gamma_B}$ and
$G^{\Gamma_C}$ (with respect to ${\cal H}_B$ and ${\cal H}_C$,
respectively) can be defined in a similar way.

For tripartite states we denote
$\Delta(G)=\Delta(G^{\Gamma_A})=\Delta(G^{\Gamma_B})=\Delta(G^{\Gamma_C})$
as the {\it degree condition}. Hildebrand {\it et
al.}$^{\tiny\cite{roland}}$ proved that the degree criterion is
equivalent to PPT criterion.  It is easy to show that this
equivalent condition is still true for the tripartite states. Thus
from Peres criterion we can get:

{\bf Theorem 3}\quad Let $\rho(G)$ be the density matrix of a graph
on $n=mpq$ vertices. If $\rho(G)$ is separable in $C^m_A\otimes
C^p_B\otimes C^q_C$, then
$\Delta(G)=\Delta(G^{\Gamma_A})=\Delta(G^{\Gamma_B})=\Delta(G^{\Gamma_C}).$

Let $G$ be a graph on $n=mpq$ vertices: $\alpha_1,\ \alpha_2,\
\cdots,\ \alpha_n$ and $f$ edges:
$\{\alpha_{i_1},\
\alpha_{j_1}\}$,\ $\{\alpha_{i_2},\ \alpha_{j_2}\},$ $\cdots$,\
$\{\alpha_{i_f}$,\ $\alpha_{j_f}\}.$ Let vertices $\alpha_s=u_iv_jw_k,$
where $s=(i-1)pq+(j-1)q+k, 1\leq i\leq m,\ 1\leq j\leq p,\ 1\leq
k\leq q.$ The vectors $|u_i\ra's,\ |v_j\ra's,\ |w_k\ra's$ form
orthonormal bases of $C^m,\ C^p$ and $C^q$, respectively. The edge
$\{u_iv_jw_k,\ u_rv_sw_t\}$ is said to be {\it entangled} if $i\neq
r,\ j\neq s,\ k\neq t.$

Consider a cuboid with $mpq$ points whose length is $m$, width is
$p$ and height is $q$, such that the distance between two
neighboring points on the same line is 1. A {\it nearest point
graph} is a graph whose vertices are identified with the points of
the cuboid and the edges have length 1, $\sqrt{2}$ and $\sqrt{3}.$

The degree condition is still a sufficient condition of the
tripartite separability for the density matrix of a nearest point
graph.

{\bf Theorem 4}\quad Let $G$ be a nearest point graph on $n=mpq$
vertices. If
$\Delta(G)=\Delta(G^{\Gamma_A})=\Delta(G^{\Gamma_B})=\Delta(G^{\Gamma_C})$,
then the density matrix $\rho(G)$ is tripartite separable in
$C^m_A\otimes C^p_B\otimes C^q_C.$

{\bf Proof. }\quad Let $G$ be a nearest point graph on $n=mpq$
vertices and $f$ edges. We associate to $G$ the orthonormal basis
$\{|\alpha_{l}\ra:l=1,\ 2,\ \cdots,\ n\} =\{|u_{i}\ra\otimes
|v_{j}\ra\otimes |w_{k}\ra:\ i=1,\ 2,\ \cdots, m;\ j=1,\ 2,\
\cdots,\ p;\ k=1,\ 2,\ \cdots,\ q\},$
where
$\{|u_{i}\ra:\ i=1,\ 2,\
\cdots,\ m\}$ is an orthonormal basis of $C^m_A$, $\{|v_{j}\ra:\
j=1,\ 2,\ \cdots,\ p\}$ is an orthonormal basis of $C^p_B$ and
$\{|w_{k}\ra:\ i=1,\ 2,\ \cdots,\ q\}$ is an orthonormal basis of
$C^q_C$.
Let
$i,\ r\in\{1,\ 2,\ \cdots,\ m\},\ j,\ s\in\{1,\ 2,\
\cdots,\ p\},\ k,\ t\in\{1,\ 2, \ \cdots,\ q\},\ \lambda_{ijk,\
rst}\in \{0,\ 1\}$
be defined by
$$\lambda_{ijk,\ rst}=\left\{
\begin{array}{ll}
1, & \hbox { if\ $(u_{i}v_{j}w_{k},\ u_{r}v_{s}w_{t})\in E(G);$}\\
0, & \hbox { if\ $(u_{i}v_{j}w_{k},\ u_{r}v_{s}w_{t})\notin E(G),$}
\end{array}
\right.$$
where $i,\ j,\ k,\ r,\ s,\ t$ satisfy either of the
following seven conditions:
\begin{itemize}
\item $i=r,\ j=s,\ k=t+1;$
\item $i=r,\ j=s+1,\ k=t;$
\item $i=r+1,\ j=s,\ k=t;$
\item $\ i=r,\ j=s+1,\ k=t+1;$
\item $i=r+1,\ j=s+1,k=t;$
\item $i=r+1,\ j=s,\ k=t+1;$
\item $i=r+1,\ j=s+1,\ k=t+1.$
\end{itemize}

Let $\rho(G),\ \rho(G^{\Gamma_A}),\ \rho(G^{\Gamma_B})$ and
$\rho(G^{\Gamma_C})$ be the density matrices corresponding to the
graph $G,\ G^{\Gamma_A},\ G^{\Gamma_B}$ and $G^{\Gamma_C}$,
respectively.
Thus
$$\begin{array}{rll}
\rho(G)&=\frac{1}{2f}(\Delta(G)-M(G)),\
&\rho(G^{\Gamma_{A}})=\frac{1}{2f}
(\Delta(G^{\Gamma_{A}})-M(G^{\Gamma_{A}})),\\[5mm]
\rho(G^{\Gamma_{B}})&=\frac{1}{2f}
(\Delta(G^{\Gamma_{B}})-M(G^{\Gamma_{B}})),\
&\rho(G^{\Gamma_{C}})=\frac{1}{2f}
(\Delta(G^{\Gamma_{C}})-M(G^{\Gamma_{C}})).
\end{array}$$
Let $G_1$ be the subgraph of $G$ whose edges are all the entangled
edges of $G$. An edge $\{u_iv_jw_k,\ u_rv_sw_t\}$ is entangled if
$i\neq r,\ j\neq s,\ k\neq t$.
Let $G^A_1$ be the subgraph of
$G^{\Gamma_A}$ corresponding to all the entangled edges of
$G^{\Gamma_A},\  G^B_1$ be the subgraph of $G^{\Gamma_B}$
corresponding to all the entangled edges of $G^{\Gamma_B}$, and
$G^C_1$ be the subgraph of $G^{\Gamma_C}$ corresponding to all the
entangled edges of $G^{\Gamma_C}.$
Obviously,
$G^A_1=(G_1)^{\Gamma_A},\ G^B_1=(G_1)^{\Gamma_B},\
G^C_1=(G_1)^{\Gamma_C}.$ We have
$$\rho(G_1)=\frac{1}{f}\sum_{i=1}^m\sum_{j=1}^p\sum_{k=1}^q
\lambda_{ijk,\
rst}P[\frac{1}{\sqrt{2}}(|u_iv_jw_k\ra-|u_rv_sw_t\ra)],$$
where
$i,\
j,\ k;\ r,\ s,\ t$ must satisfy either of the above seven
conditions. We can get $\rho(G^A_1),\ \rho(G^B_1)$ and $\rho(G^C_1)$
by commuting the index of $u,\ v,\ w$ in the above equation,
respectively.
Also we have
$$\Delta(G_1)=\frac{1}{2f}\sum_{i=1}^m\sum_{j=1}^p\sum_{k=1}^q
\lambda_{ijk,\ rst}P[|u_iv_jw_k\ra],$$ where $i,\ j,\ k;\ r,\ s,\ t$
must satisfy either of the above seven conditions. We can get
$\Delta(G^A_1),\ \Delta(G^B_1)$ and $\Delta(G^C_1)$ by commuting the
index of $\lambda$ with respect to the Hilbert space ${\cal H}_A,\
{\cal H}_B,\ {\cal H}_C$, respectively. Let $G_2,\ G^A_2,\ G^B_2$
and $G^C_2$ be the subgraph of $G,\ G^A,\ G^B$ and $G^C$ containing
all the unentangled edges, respectively.
It is obvious that
$\Delta(G_2)=\Delta(G^{\Gamma_A}_2)=\Delta(G^{\Gamma_B}_2)
=\Delta(G^{\Gamma_C}_2).$ So
$\Delta(G)=\Delta(G^{\Gamma_A})=\Delta(G^{\Gamma_B})
=\Delta(G^{\Gamma_C})$
if and only if
$\Delta(G_1)=\Delta(G^{\Gamma_A}_1)=\Delta(G^{\Gamma_B}_1)
=\Delta(G^{\Gamma_C}_1).$
The degree condition implies that
$$\lambda_{ijk,\ rst}=\lambda_{rjk,\
ist}=\lambda_{isk,\ rjt}=\lambda_{ijt ,\ rsk},$$
for any $i,\
r\in\{1,\ 2,\ \cdots,\ m\},\ j,\ s \in\{1,\ 2,\ \cdots,\ p\},\ \ k,\
t \in\{1,\ 2,\ \cdots,\ q \}.$
The above equation shows that
whenever there is an entangled edge $\{u_{i}v_{j}w_{k},\
u_{r}v_{s}w_{t}\}$ in $G$ (here we must have $i\neq r,\ j\neq s,\
k\neq t$), there must be the entangled edges $\{u_{r}v_{j}w_{k},\
u_{i}v_{s}w_{t}\},\ \{u_{i}v_{s}w_{k},\ u_{r}v_{j}w_{t}\}$
and
$\{u_{i}v_{j}w_{t},\ u_{r}v_{s}w_{k}\}$ in $G$.
Let
$$\begin{array}{rl}
\rho(i,\ j,\ k;\ r,\ s,\ t) =&\frac{1}{4}(P[\frac{1}{\sqrt{2}}
(|u_{i}v_{j}w_{k}\ra-|u_{r}v_{s}w_{t}\ra)]
+P[\frac{1}{\sqrt{2}}(|u_{r}v_{j}w_{k}\ra-|u_{i}v_{s}w_{t}\ra)]\\[5mm]
&+P[\frac{1}{\sqrt{2}}(|u_{i}v_{s}w_{k}\ra-|u_{r}v_{j}w_{t}\ra)]
+P[\frac{1}{\sqrt{2}}(|u_{i}v_{j}w_{t}\ra-|u_{r}v_{s}w_{k}\ra)]).
\end{array}$$
By Lemma 2, we know $\rho(i,\ j,\ k;\ r,\ s,\ t)$ is tripartite
separable in $C^m_A\otimes C^p_B\otimes C^q_C.$
By Theorem 3 in
\cite{sam2} we can easily get $\rho(G_2)$ is tripartite separable in
$C^m_A\otimes C^p_B\otimes C^q_C.$ $\Box$

From Theorems 3 and 4 we can obtain the following corollary which is
a sufficient and necessary criterion (we called {\it
degree-criterion}) of the density matrix of a nearest point graph:

{\bf Corollary 1}\quad Let $G$ be a nearest point graph on $n=mpq$
vertices, then the density matrix $\rho(G)$ is tripartite separable
in $C^m_A\otimes C^p_B\otimes C^q_C$ if and only if
$\Delta(G)=\Delta(G^{\Gamma_A})=\Delta(G^{\Gamma_B})=\Delta(G^{\Gamma_C}).$

{\bf Example }\quad Let $G$ be a graph on $12=3\times 2\times 2$
vertices, having a unique edge $\{u_1v_1w_1,\ u_2v_2w_2\}$.
Then we
have
$$\rho(G)=\frac{1}{2} \left(
\begin{array}{cccccccccccc}
1&0&0&0&0&0&0&-1&0&0&0&0\\
0&0&0&0&0&0&0&0&0&0&0&0\\
0&0&0&0&0&0&0&0&0&0&0&0\\
0&0&0&0&0&0&0&0&0&0&0&0\\
0&0&0&0&0&0&0&0&0&0&0&0\\
0&0&0&0&0&0&0&0&0&0&0&0\\
0&0&0&0&0&0&0&0&0&0&0&0\\
-1&0&0&0&0&0&0&1&0&0&0&0\\
0&0&0&0&0&0&0&0&0&0&0&0\\
0&0&0&0&0&0&0&0&0&0&0&0\\
0&0&0&0&0&0&0&0&0&0&0&0\\
0&0&0&0&0&0&0&0&0&0&0&0
\end{array}
\right).$$
The partially transposed graph $G^{\Gamma_A}$ is a graph
on 12 vertices and has an edge $\{u_2v_1w_1$,\ $u_1v_2w_2\}$.
Then
$$\rho(G^{\Gamma_{A}})=\frac{1}{2} \left(
\begin{array}{cccccccccccc}
0&0&0&0&0&0&0&0&0&0&0&0\\
0&0&0&0&0&0&0&0&0&0&0&0\\
0&0&0&0&0&0&0&0&0&0&0&0\\
0&0&0&1&-1&0&0&0&0&0&0&0\\
0&0&0&-1&1&0&0&0&0&0&0&0\\
0&0&0&0&0&0&0&0&0&0&0&0\\
0&0&0&0&0&0&0&0&0&0&0&0\\
0&0&0&0&0&0&0&0&0&0&0&0\\
0&0&0&0&0&0&0&0&0&0&0&0\\
0&0&0&0&0&0&0&0&0&0&0&0\\
0&0&0&0&0&0&0&0&0&0&0&0\\
0&0&0&0&0&0&0&0&0&0&0&0
\end{array}
\right).$$ Obviously, the degree matrices of $G$ and $G^{\Gamma_A}$
are different. The eigenvalues of $\rho(G)^{T_A}$ are 0 (with
multiplicity 8), $\frac{1}{2}$ (with multiplicity 3) and
$-\frac{1}{2}$, so $\rho(G)^{T_A}$ is not positive semidefinite.
According to Peres criterion, $\rho(G)$ is tripartite entangled.
$\Box$

Two graphs $G$ and $H$ are said to be {\it isomorphic}, denoted as
$G\cong H$, if there is an isomorphism between $V(G)$ and $V(H)$,
i.e., there is a permutation matrix $P$ such that
$PM(G)P^T=M(H).^{\tiny\cite{sam1}}$

{\bf Theorem 5}\quad Let $G$ and $H$ be two graphs on $n=mpq$
vertices. If $\rho(G)$ is tripartite entangled in $C^m\otimes
C^p\otimes C^q$ and $G\cong H$, then $\rho(H)$ is not necessarily
tripartite entangled in $C^m\otimes C^p\otimes C^q.$

{\bf Proof.}\quad Let $G$ be the graph introduced in the above
example.
Then $\rho(G)$ is tripartite entangled.
Let $H$ be a graph
on 12 vertices, having an edge $\{u_1v_1w_1,\ u_1v_1w_2\}$.
Obviously, $G$ is isomorphic to $H$. However,
$$\rho(H)=P[\frac{1}{\sqrt{2}}
(|u_1v_1w_1\ra-|u_1v_1w_2\ra)]=|u_1\ra\la u_1|\otimes|v_1\ra\la
v_1|\otimes|w^+\ra\la w^+|,$$
where
$|w^+\ra=\displaystyle\frac{1}{\sqrt{2}}(|w_1\ra-|w_2\ra),$ shows
that $\rho(H)$ is tripartite separable. $\Box$

\end{document}